# Efficient fidelity control by stepwise nucleotide selection in polymerase elongation

2/13/14 11:35 PM Jin Yu[1]
[1]Beijing Computational Science Research Center
No. 3 Heqing Road, Haidian District, Beijing, China 100084
E-mail address: jinyu@csrc.ac.cn

Polymerases select nucleotides before incorporating them for chemical synthesis during gene replication or transcription. How the selection proceeds stepwise efficiently to achieve sufficiently high fidelity and speed is essential for polymerase function. We examined step-by-step selections that have conformational transition rates tuned one at time in the polymerase elongation cycle, with a controlled differentiation free energy at each checkpoint. The elongation is sustained at non-equilibrium steady state with constant free energy input and heat dissipation. It is found that error reduction capability does not improve for selection checkpoints down the reaction path. Hence, it is essential to select early to achieve an efficient fidelity control. In particular, for two consecutive selections that reject the wrong substrate back and inhibit it forward from a same kinetic state, the same error rates are obtained at the same free energy differentiation. The initial screening is indispensible for maintaining the elongation speed high, as the wrong nucleotides can be removed quickly and replaced by the right nucleotides at the entry. Overall, the elongation error rate can be repeatedly reduced through multiple selection checkpoints. The study provides a theoretical framework to conduct further detailed researches, and assists engineering and redesign of related enzymes.

## I. INTRODUCTION

Polymerases are essential enzymes in charge of gene replication or transcription [1]. A polymerase moves along DNA or RNA while synthesizing a new strand of nucleic acid, largely according to Watson-Crick base pairing with the template strand. Without the polymerase, the template-based polymerization can also happen, but at an extremely low speed and with low fidelity. The polymerase essentially catalyzes the polymerization and improves the fidelity. Being a nanometer-sized molecular machine, the polymerase works under high viscosity and significant thermal noises. How to achieve sufficiently high fidelity at a sufficiently high elongation speed is thus key to the polymerase function.

From what had been measured experimentally, it was suggested that the polymerase moves as a Brownian ratchet along the DNA/RNA track [2-7]. Upon binding and insertion of the incoming nucleotide, backward translocation of the polymerase is inhibited. The nucleotide insertion often accompanies with substantial conformational changes of the polymerase [8, 9]. Following the insertion, the nucleotide covalently links to the newly synthesized chain through phosphoryl transfer reaction (see **Fig 1a**). The catalytic reaction is followed by pyrophosphate ion (PPi) release, which concludes the enzymatic cycle, so that the polymerase can translocate forward and recruit the next nucleotide. From the recruitment to the end of the catalysis, the nucleotide can be selected at multiple kinetic checkpoints.

A wrong nucleotide or an error can be selected against by the polymerase, via destabilizing the intermediate state to increase the backward transition rate (*rejection*), or via raising the forward activation barrier to decrease the forward transition rate (*inhibition*) along the reaction path. The nucleotide selection is common to fidelity control of all polymerases. The error rate achieved by the selection can reach as low as one in tens of thousands to one in a million ($10^{-4} \sim 10^{-6}$) [10]. After the catalysis, or once the nucleotide is covalently added, the error can be further corrected through exo- or endo-nuclease reaction. The enzymatic reaction excises wrong nucleotides, serving for proofreading. In general, the fidelity of the polymerization is controlled through both the nucleotide selection and proofreading [8, 9, 11-14]. The error rate can be lowered by one to three orders of magnitudes further by the proofreading, to as low as one in tens of millions or even lower [10].

The kinetic proofreading had been widely discussed in the context of genetic control [15-19]. In order to achieve high specificity or fidelity, the enzyme and substrate can form multiple intermediate states before generating the final product. The intermediate states are made through driven reactions (breaking the detailed balance) with energy sources. Each kinetic proofreading procedure is implemented through a branching or looping reaction that breaks one of the intermediates back into the *apo* enzyme and substrate (see **Fig 1b**). The free energy for differentiation between the right and wrong substrates can thus be repeatedly utilized at those intermediates to achieve high fidelity, through a cascade of the proofreading steps. The proofreading related activities of the polymerases have also been detected at single molecule level in recent years [3, 20, 21], which inspired further modeling studies [22-25].

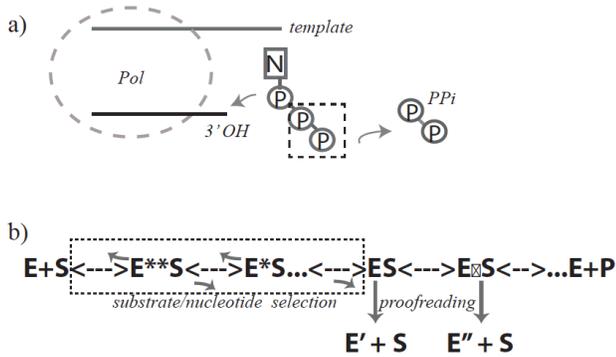

FIG. 1. Schematics of polymerase reaction and stepwise nucleotide selection. **(a)** Polymerase (*Pol*) enzyme (**E**) catalyzes a phosphoryl transfer reaction E*NA$_n$+ NTP ↔ E*NA$_{n+1}$+ PPi. The incoming NTP is incorporated according to the template strand. **(b)** The kinetic scheme for incorporating the nucleotide substrate during polymerase elongation. Nucleotide or substrate selection can happen at any checkpoint prior to chemical catalysis that leads to formation of **ES**, through backward *rejection* or forward *inhibition* (see text). Proofreading happens after formation of **ES**, while before the final product formation (**E+P**). Each proofreading step throws errors away via a branched driven reaction, for example, the one destroying **ES** back into **E´+S**.

The proofreading-free selection, however, has attracted less attention as it appears simple. Early work suggested that the error ratio of the selection cannot be lower than exp(-$\Omega_{max}$/$k_B T$), which is determined by a maximum free energy differentiation $\Omega_{max}$ between the right (cognate) and wrong (non-cognate) substrate along the reaction path [17]. Individual steps of the selection had not been further considered. The overall characterization seemed to suffice as if the selection details were not accessible. Nevertheless, stepwise mechanism of transcription fidelity has been revealed recently, for example, from the structure-based mutagenesis and kinetic analysis [26]. Individual steps of the substrate selection have been characterized to contribute to the overall fidelity. On the other hand, modeling and computation technologies allow protein structural dynamics to be captured down to atomistic scale, providing opportunities to characterize detailed selection mechanisms. Here we present a model framework to study the stepwise substrate selection, in particular, the nucleotide selection during the polymerase elongation. The selection relies on multiple intermediate states prior to the end of the catalysis. Either the backward transition to the previous state is enhanced, when the intermediate structure is bound with the wrong substrate, or the transition toward the next state is inhibited. Each modulated transition constitutes an "elementary selection" addressed below. The elementary selection happens between two consecutive states along the reaction path (see **Fig 1b**), *without* branching back to the *apo* state as that in the proofreading. The selection is sustained at non-equilibrium steady state (NESS), as long as the polymerase elongates at a nonzero speed [27, 28].

The key question we want to address in this study is how to conduct stepwise selection efficiently during the polymerase elongation. Being 'efficient' here means to achieve a sufficiently low error rate at a sufficiently high speed, when the free energy differentiation is limited and controlled. With *given* kinetic parameters for incorporating the *right* substrates, and $\Omega_{max}$ as a control parameter, we wanted to find comparatively 'efficient' selection strategies, or parameter sets for the *wrong* substrate kinetics that lower the error rate without necessarily lowering much the speed. Indeed, one can dissect $\Omega_{max}$ into individual terms $\{\Delta_i^{\pm}\}$ (with $\sum \Delta_i^{\pm} = \Omega_{max}$) at the multiple selection checkpoints. One can see in this work how the elongation speed and error rate vary as $\{\Delta_i^{\pm}\}$ are allocated differently among the checkpoints along the reaction path.

In early studies, an 'efficiency-accuracy' tradeoff was discovered in substrate selection [19, 29, 30]. The tradeoff means that a selection system operates close to its maximal accuracy when the enzyme efficiency approaches to zero. The maximal accuracy is determined by exp($\Omega_{max}$/$k_B T$), and both the accuracy and enzyme efficiency vary depending on the kinetic rates of the system. The tradeoff shows a limit of the selection as the system kinetics parameters vary under experimentally designed conditions, and helps to extract kinetic information of the system [19]. In current study, however, we consider how the error rate and speed vary among different selection strategies, without varying kinetics for the right substrate incorporation.

In this work, we adopted polymerase elongation schemes in general, while using data from T7 RNA polymerase (RNAP) [4, 7, 31] to demonstrate numerical results. T7 RNAP elongates at an error rate ~ 10$^{-4}$ without proofreading activities detected [32]. It is an ideal system to study the nucleotide selection. We analyze first a generic three-state kinetic scheme, building connections with previous quantitative studies. The basic findings are re-examined in a more specific elongation scheme with five states. The kinetic schemes apply to most of polymerases, though different rate-limiting steps happen in different cases. Accordingly, how the findings vary as the rate-limiting step varies is also addressed. In addition, implementations of the present framework to exemplary polymerases are introduced as well. The entropy production and heat dissipation during the information acquisition process of elongation are discussed in the end.

## II. ELEMENTARY SELECTIONS AND SELECTION STRENGTH

For the template-based nucleotide incorporation, we consider that the polymerases recognize the nucleotides either as right or wrong. Below we use free energy profiles for incorporating both the right and wrong nucleotides to characterize the stepwise selection, starting from the nucleotide binding to the end of the catalytic reaction. Independent of enzymatic activities, the free energy input (>0) upon incorporating a right nucleotide and a wrong one is $\Delta G_c^r$ and $\Delta G_c^w$, respectively. The overall difference is $\Delta G_c^r - \Delta G_c^w = \delta_G$ (≥0). When there is no enzyme, the maximum free energy differentiation $\Omega_{max} = \delta_G$. The enzyme activities modulate intermediate stabilities or barriers along the reaction path to greatly accelerate the right substrate incorporation while deter the wrong substrate incorporation. Consequently, $\Omega_{max}$ rises largely above $\delta_G$, while $\Delta G_c^r$, $\Delta G_c^w$ and $\delta_G$ keep unchanged.

First, we consider *elementary selections* that tune the free energy profile of incorporating the *wrong* substrates, for only one transition/activation barrier at a time. **Fig 2a** shows the free energy profiles of the wrong substrate incorporation (dashed line) for typical elementary selections along the reaction path, on top of the profile of the right substrate incorporation (solid line). The leftmost one, denoted $S_i^-$, selects against the wrong substrates by destabilizing state *i* or lowering the backward transition barrier from *i* to *i-1* (not shown) by $\Delta_i^- = \Delta$ (>0), in comparison with that of the right substrate; the free energy profile thereafter does *not* distinguish between the right and wrong, except for a difference $\delta_G$ in the end. Consequently, the backward transition rate from *i* to *i*-1 becomes larger for the wrong than that for the right. One can characterize the selection strength as the ratio between the backward rates ($k_{i-}$) of the wrong and right $\eta_i^- = k_{i-}^w / k_{i-}^r = e^{\Delta/k_BT}$, with '*w*' labeling for the wrong and '*r*' labeling for the right. A strong selection against the wrong substrate corresponds to a large $\eta$ (>>1). The selection denoted $S_i^+$ next down the reaction path, however, inhibits the wrong nucleotides from transiting from state *i* to *i+1*, by raising the forward free energy barrier by $\Delta_i^+ = \Delta$ (but not to the backward barrier from *i+1* to *i*). Correspondingly, the selection strength is defined as the ratio between the forward rates ($k_{i+}$) of the right and wrong $\eta_i^+ = k_{i+}^r / k_{i+}^w = e^{\Delta/k_BT}$. Following $S_i^+$ and similar to $S_i^-$, the selection denoted $S_{i+1}^-$ (with a strength $\eta_{i+1}^-$) selects against wrong nucleotides by reducing the backward transition barrier from *i+1* to *i* by $\Delta$. A same differentiation free energy $\Delta$ is used for different elementary selections to facilitate comparing their performances.

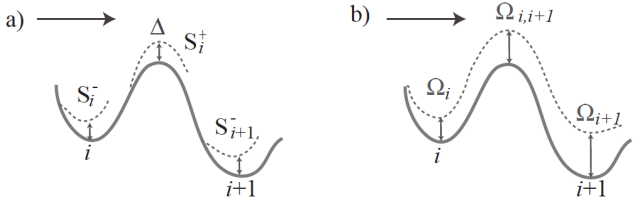

FIG. 2. Elementary selections along the reaction path. **(a)** Typical elementary selections. The solid line represents the free energy profile of incorporating the right substrates, while the dashed line shows that for the wrong substrates under the selection. Each elementary selection changes the kinetic rate of a backward or forward transition for incorporating the wrong. Selection $S_i^-$ and $S_{i+1}^-$ work through *rejection* while selection $S_i^+$ works through *inhibition*. **(b)** A selection as a combination of $S_i^-$ (strength $\eta_i^- = e^{\Omega_i/k_BT}$), $S_i^+$ ($\eta_i^+ = e^{(\Omega_{i,i+1}-\Omega_i)/k_BT}$), and $S_{i+1}^-$ ($\eta_{i+1}^- = e^{(\Omega_{i+1}-\Omega_{i,i+1})/k_BT}$).

To keep the overall free energy difference between the right and wrong substrate incorporation $\Delta G_c^r - \Delta G_c^w = \delta_G$ unchanged, a 'reset' of the free energy difference to $\delta_G$ (> 0) for each selection is implemented at the end of the catalysis (see illustrations later).

By defining the elementary selections along the reaction path, one can construct any substrate selection strategy as a combination of those elementary selections. For example, in **Fig 2b**, a selection has the free energy profile for incorporating the wrong substrate raised stepwise by $\Omega_i$, $\Omega_{i,i+1}$ and $\Omega_{i+1}$ (at state *i*, the intermediate between *i* and *i+1*, and state *i+1*) above that of incorporating the right nucleotide. The selection can be regarded as a combination of elementary selections $S_i^-$ with the differentiation free energy $\Delta_i^- = \Omega_i$, $S_i^+$ with $\Delta_i^+ = \Omega_{i,i+1} - \Omega_i$, and $S_{i+1}^-$ with $\Delta_{i+1}^- = \Omega_{i+1} - \Omega_{i,i+1}$. Hence, $\Omega_{i,i+1} = \Delta_i^- + \Delta_i^+$ and $\Omega_{i+1} = \Delta_i^- + \Delta_i^+ + \Delta_{i+1}^-$ (in this case $\Omega_{max} = \Omega_{i+1}$ as $\Delta_i^\pm \geq 0$) indeed measure the *accumulative* free energy differentiation. For an efficient selection, it is necessary $\Delta_i^\pm \geq 0$ so that the stepwise free energy difference $\Omega_i$ grows along the reaction path. Below, we focus on how the *elementary selections* impact on the polymerization speed and error rate, since any selection in general applies as the elementary selections combined together.

## III. THREE-STATE ELONGATION SCHEME

We start with a generic three-state kinetic scheme (**Fig 3a**) to compare two basic nucleotide selection strategies in the elongation cycle, the *rejection* and the *inhibition*. The scheme consists of the pre-translocated (*I*), post-translocated (*II*), and substrate state (*III*). Upon translocation (*I→II*), an incoming NTP diffuses and binds to the polymerase (*II→III*), prior to being recognized as right or wrong. A catalytic step then follows (*III→I*). Correspondingly, recognition and selection of the nucleotide happen either upon substrate binding (*III→II*) though the rejection, or at the catalytic stage (*III→I*) through the inhibition. **Fig 3b** illustrates schematically the selection strategies on the free energy profile, with solid and dashed lines for incorporating the right and wrong species, respectively.

Under the initial rejection $S_{III}^-$, the unbinding or off-rate of the wrong nucleotide ($k_{III-}^w$) becomes larger than that of the right nucleotide ($k_{III-}^r$). One can quantify the selection strength as $\eta_{III}^- = k_{III-}^w / k_{III-}^r = e^{\Delta/k_B T}$, with $\Delta$ (>0) measuring the free energy difference between the wrong and right species detected at this checkpoint (*III→II*). As mentioned, one resets $\delta_G$ at the end of the cycle (see **Figure 3b**).

Alternatively, the catalytic inhibition $S_{III}^+$ raises the activation barrier for the catalysis (*III→I*) of the wrong nucleotide above that of the right. One can quantify the selection strength by $\eta_{III}^+ = k_{III+}^r / k_{III+}^w = e^{\Delta/k_B T}$. Similarly, one resets $\delta_G$ in the end.

To consider probability fluxes for both the right and wrong species in the three-state scheme, one can define a population vector $\Pi = (P_I, P_{II}, P_{III}^r, P_{III}^w)^T$ to represent the probability distributions of states *I*, *II* and *III* (for both right and wrong species). The master equation is:

$$\tfrac{d}{dt}\Pi = M\Pi \qquad (1)$$

where *M* is a 4x4 transition matrix as

$$\begin{bmatrix} -(1-Err)k_{I-} - Err \cdot k_{I-}\dfrac{\eta_G}{\eta_{III}^- \eta_{III}^+} - k_{I+} & k_{II-} & k_{III+} & \dfrac{k_{III+}}{\eta_{III}^+} \\ k_{I+} & -k_{II-}-k_{II+} & k_{III-} & k_{III-}\eta_{III}^- \\ (1-Err)k_{I-} & i_r k_{II+} & -k_{III-}-k_{III+} & 0 \\ Err \cdot k_{I-}\dfrac{\eta_G}{\eta_{III}^- \eta_{III}^+} & (1-i_r)k_{II+} & 0 & -k_{III-}\eta_{III}^- - \dfrac{k_{III+}}{\eta_{III}^+} \end{bmatrix}$$

with $k_{I+}$ and $k_{II-}$ the forward and backward translocation rate, $k_{II+}$ and $k_{III-}$ the binding ($\propto [NTP]$) and unbinding rate of the nucleotide, and $k_{III+}$ and $k_{I-}$ the catalytic and its reverse rate. $i_r$ is the portion of right nucleotides from solution at 'input' ($i_r = 1/4$ by default for four equally mixed nucleotides in solution). *Err* is the 'output' or elongation error rate at the end of the cycle, after nucleotide selection. $\eta_G \equiv e^{\delta_G}$ is to keep the overall free energy difference between the right and wrong nucleotide incorporation to $\delta_G$.

Using the steady state solution for Eq (1), at $k_{I-} \to 0$ for simplicity, one obtains the probability flux or the polymerization/elongation rate $J$ (or the speed $v = l_0 J$ with $l_0$ =1bp),

$$J = \dfrac{\dfrac{k_{\max}^0}{1+\Gamma}[NTP]_{total}}{\dfrac{\Lambda}{1+\Gamma}K_M^0 + [NTP]_{total}} \qquad (2)$$

where $k_{\max}^0$ and $K_M^0$ are the maximal rate and Michaelis constant when there is *no* nucleotide selection ($\eta_{III}^- = \eta_{III}^+ = 1$). The modulation constants $\Gamma$ and $\Lambda$ depend on the selection strength $\eta_{III}^-$ and $\eta_{III}^+$. The expressions of the constants can be found in **Supplementary Material (SM), Appendix I**.

### III.1 Speed modulation by the selection

When there is no selection, $J = J_0 \equiv \dfrac{k_{\max}^0 [NTP]_{total}}{K_M^0 + [NTP]_{total}}$ ($\Gamma \sim 0$ and $\Lambda = 1$). Below we show how the selections affect the polymerization rates for three typical cases.

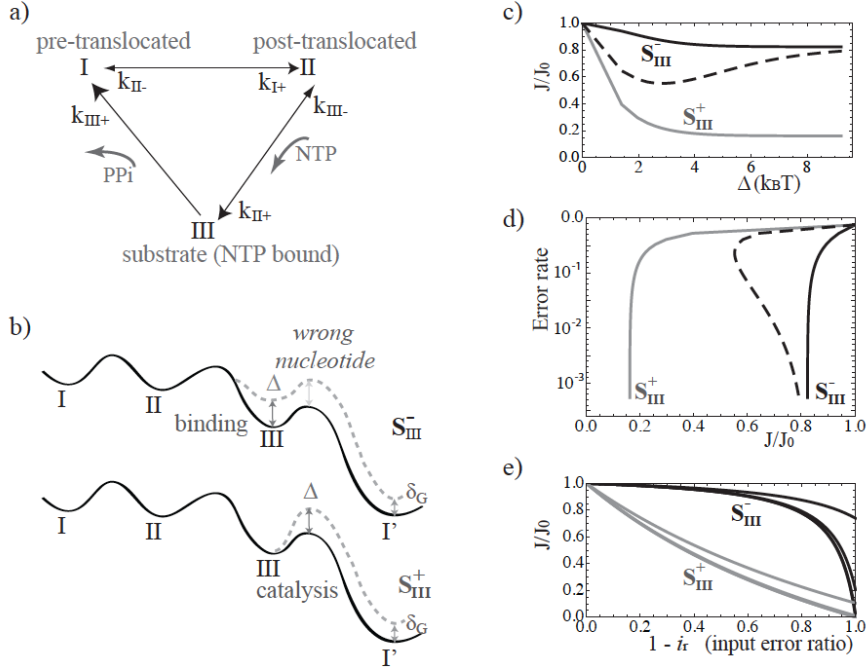

FIG. 3. Nucleotide selections in the three-state elongation scheme. **(a)** The three-state scheme consists of translocation, NTP binding, and catalysis / PPi release. **(b)** Elementary selections demonstrated on the free energy profile (incorporating the right/wrong nucleotide in solid /dashed line). The initial screening or rejection is through $S_{III}^-$, while $S_{III}^+$ selects through catalytic inhibition. $\Delta G_c^r - \Delta G_c^w = \delta_G$ is set in the end. **(c)** The polymerization or elongation rate vs. the differentiation free energy $\Delta$ for respective selections $S_{III}^-$ (dark line) and $S_{III}^+$ (gray). The elongation rate is normalized as $J/J_0$, with $J_0$ the elongation rate without the selection. A combined selection strategy with $S_{III}^-$ and $S_{III}^+$ at equal strength $\eta = e^{\Delta/2k_BT}$ is also shown (dashed line). **(d)** The error rate vs. the elongation rate under $S_{III}^-$, $S_{III}^+$, and the combined strategy, as $\Delta$ varies. **(e)** The elongation rate vs. the input error ratio. The input error ratio is measured by $1-i_r$, with $i_r$ the portion of right nucleotides from solution. For selections $S_{III}^-$ and $S_{III}^+$, lines are shown respectively as black and gray at different selection strengths (for $\eta$ = 10, $10^2$, $10^3$, and $10^4$, from up to down).

**i)** When *only* the initial rejection $S_{III}^-$ works ($\eta_{III}^- > 1, \eta_{III}^+ = 1$). In this case, $\Gamma = 0$; as $\eta_{III}^-$ increases, $\Lambda$ grows from 1 to $1/i_r$, and $J \to J_1 \equiv \dfrac{k_{max}^0 [NTP]_{right}}{K_M^0 + [NTP]_{right}}$ with $[NTP]_{right} = i_r [NTP]_{total}$. That says, when strong selection exists through NTP unbinding, the polymerization rate converges to that for a single species of right nucleotides. Usually $J_1 < J_0$ as $[NTP]_{right} < [NTP]_{total}$. When the NTP concentration is sufficiently high ($[NTP]_{right} > K_M^0$), $J_1 \to k_{max}^0$ and $J_1 \sim J_0$, so that the speed keeps high.

**ii)** When *only* the catalytic inhibition $S_{III}^+$ works ($\eta_{III}^- = 1, \eta_{III}^+ > 1$). The catalytic rate is effectively reduced to $k_{max}^0/(1+\Gamma)$ as $\Gamma > 0$. When $\eta_{III}^+$ increases, $J \to \dfrac{k_{max}^0 [NTP]_{right}}{K_M^0 + (1+\Gamma)[NTP]_{right}} < J_1$. At a high NTP concentration, $J \to k_{max}^0/(1+\Gamma)$, the rate is significantly reduced below $J_0$ as long as the translocation is not rate limiting. In case that the translocation happens much slower than the catalytic step, $\Gamma \to 0$ holds (see **SM Appendix I**), then the saturating elongation rate can still keep high under the selection.

In **Fig 3c**, we show the relative speed ($J/J_0$) vs. the error-detection energy ($\Delta$) under selections $S_{III}^-$ and $S_{III}^+$, respectively. By default, translocation is set fast as that was reported or commonly assumed [3, 4]. Accordingly, one sees that the strong nucleotide selection through the catalytic inhibition lowers the polymerization rate significantly.

**iii)** When *both* selections $S_{III}^-$ and $S_{III}^+$ work ($\eta_{III}^- > 1$ and $\eta_{III}^+ > 1$). As both selections get strong, $\Gamma \to 0$, $\Lambda \to 1/i_r$, hence, $J \to \dfrac{k_{max}^0 [NTP]_{right}}{K_M^0 + [NTP]_{right}} = J_1$. That says, under a *combined* selection strategy, the elongation rate can still keep high, approaching to that under $S_{III}^-$ alone. In **Fig 3c** the elongation rate decreases a bit for $\Delta < 3$ k$_B$T but then *increases* for $\Delta > 3$ k$_B$T to approach $J_1$. The combined selection strategy here relies equally on $S_{III}^-$ and $S_{III}^+$, with the same differentiation free energy $\Delta/2$ for each. Hence, even a small free energy differentiation (1~2 k$_B$T) at the initial screening can keep the relative speed ($J/J_0$) above 0.5. Indeed, the higher contribution from $S_{III}^-$ in the combined strategy, the faster the polymerization rate converges to $J_1$.

### III.2 Error reduction by the selection

In order to compare the polymerization rates for the right and wrong nucleotides as they compete for binding, one writes $J = J^r + J^w$ as

$$J^r = i_r \Lambda J = \frac{\frac{\Lambda k_{max}^0}{1+\Gamma}[NTP]_{right}}{\frac{\Lambda}{1+\Gamma}K_M^0 + [NTP]_{right} + [NTP]_{wrong}} \quad (2a)$$

$$J^w = (1-i_r\Lambda)J = \frac{\frac{(1-i_r\Lambda)k_{max}^0}{(1-i_r)(1+\Gamma)}[NTP]_{wrong}}{\frac{\Lambda}{1+\Gamma}K_M^0 + [NTP]_{right} + [NTP]_{wrong}} \quad (2b)$$

As a result, one obtains the output error rate *Err*, as the polymerization rate of the wrong nucleotides over that of *both* the right and wrong:

$$Err \equiv \frac{J^w}{J} = 1 - i_r\Lambda = \frac{1-i_r}{1+i_r\frac{k_{III-}}{k_{III+}+k_{III-}}(\eta_{III}^-\eta_{III}^+ - 1)} \quad (3)$$

From **Eq 3** above, one sees $Err \sim \frac{1-i_r}{i_r}(1+\kappa)e^{-2\Delta/k_BT}$ at the strong selection limit ($\eta_{III}^- = \eta_{III}^+ = e^{\Delta/k_BT} >> 1$, $\kappa \equiv k_{III+}/k_{III-}$). That says, the error rate deceases exponentially with the *accumulate* differentiation free energy ($2\Delta$ here), as the selection gets strong. Interestingly, selections $S_{III}^-$ and $S_{III}^+$ impact *equally* on the error reduction: A same value or variation of the individual selection strength $\eta_{III}^-$ or $\eta_{III}^+$ gives a same error rate *Err* or error variation, due to the error rate dependence on $\eta_{III}^-\eta_{III}^+$. *Err* is independent of the overall free energy input as the elongation considered is under the strong non-equilibrium limit [27, 28].

### *III.3 Connection with previous work*

Current formulation can be easily linked to conventions that focus on *enzyme efficiency* $k_{max}/K_M$. In the absence of the wrong nucleotides or nucleotide selection, the efficiency is written as $\zeta_0 \equiv k_{max}^0/K_M^0$. In the presence of both the right and wrong nucleotides, the efficiency is altered to $\zeta = \zeta_0/\Lambda$ ($\Lambda > 1$), according to **Eq 2**.

From **Eq 2a** and **2b**, one also obtains the polymerase efficiencies for the right and wrong nucleotides as $\zeta^r = \zeta_0$ and $\zeta^w = \frac{1-i_r\Lambda}{(1-i_r)\Lambda}\zeta_0$, respectively. That says, the efficiency for incorporating the right nucleotide is fixed, while the efficiency for incorporating the wrong approaches zero as the nucleotide selection gets strong ($\Lambda \to 1/i_r$).

To characterize the fidelity level, one could alternatively use the *accuracy A*, defined as the efficiency of incorporating the right nucleotides over that of the wrong:

$$A \equiv \frac{\zeta^r}{\zeta^w} = \frac{(1-i_r)\Lambda}{1-i_r\Lambda} = \frac{k_{III-}\eta_{III}^-\eta_{III}^+}{k_{III+}+k_{III-}} \quad (3a)$$

Indeed, $A = \frac{(1-Err)/Err}{i_r/(1-i_r)}$ is to quantify how much the right portion of the nucleotides is at *output* relative to that at *input*. In contrast, *Err* only counts the percentile of wrong nucleotides at output. As $\eta_{III}^-\eta_{III}^+ = e^{(\Delta_{III}^-+\Delta_{III}^+)/k_BT} = e^{\Omega_{max}/k_BT}$ in this case, the efficiency-accuracy tradeoff can also be derived as (see details in **SM Appendix I**)

$$\zeta^r \propto \frac{e^{\Omega_{max}/k_BT} - A}{e^{\Omega_{max}/k_BT} - 1} \quad (3b)$$

where $\zeta^r$ is the efficiency incorporating the right nucleotide. Since $\zeta^r = \zeta_0$ (>0) with the given kinetic parameters for the right, the accuracy $A$ only approaches but cannot be equal to the maximum accuracy $e^{\Omega_{max}/k_BT}$ when the selection gets strong (as $\Delta$ and $\Omega_{max}$ increase). In current work, we focus only on how $J$ and *Err* vary as the selection becomes strong under different selection strategies. Both $\zeta$ and $A$ vary with *Err* but not $J$.

In **Fig 3d**, we show *Err* vs. $J/J_0$ for the respective selections $S_{III}^-$ and $S_{III}^+$, and for the combined selection. Basically, one sees that by increasing the selection strength, the error rate can be continuously lowered, while the elongation rates converge to different values and do not change further upon the strong selection. At high NTP concentration, $S_{III}^-$ keeps the converged speed high, while $S_{III}^+$ lowers the speed significantly. Notably, the combined selection approaches to that under $S_{III}^-$ alone when the selection becomes strong. In terms of the error reduction, these selections perform equally well. This is because the selections through the *rejection* ($S_{III}^-$) and *inhibition* ($S_{III}^+$) from the same state (*III*) give the same error rate at the same selection strength.

Experimentally, the individual selection strength can be quantified. When *purely wrong* nucleotides are supplied in the solution ($i_r = 0$), $J \to \dfrac{\dfrac{k_{max}^0}{\eta_{III}^+}[NTP]_{right}}{\eta_{III}^- K_M^0 + [NTP]_{right}}$. By measuring how $k_{max}$ and $K_M$ are modulated comparing assays of purely wrong and purely right nucleotides, one is able to determine $\eta_{III}^-$ and $\eta_{III}^+$ in the three-state elongation scheme. Besides, one can also probe which selection is used, semi-quantitatively, by mixing right and wrong nucleotides in variable portions in solution and monitoring how the polymerization rate changes. In **Fig 3e** we show $J/J_0$ vs. the input error ratio $1-i_r$. If the selection is through $S_{III}^-$ or includes $S_{III}^-$ (as in the combined selection), $J/J_0$ would be insensitive to $1-i_r$, keeping a concave shape. $J/J_0$ decreases about linearly as $1-i_r$ increases under $S_{III}^+$ only. The trends keep robust as long as the overall nucleotide concentration is not too low. The results for the fully reversible elongation kinetics ($k_L > 0$) are similar, and can be found in **SM Appendix II**.

# IV. A FIVE-STATE ELONGATION SCHEME

Next, we use a slightly more specific scheme with five states (**Fig 4a**) to describe the polymerase elongation cycle. Comparing to the three-state scheme, the essential difference is there are two instead of one kinetic steps ($II \to III$ and $III \to IV$) proceeding to the chemical catalysis. In T7 RNAP and some of other polymerases, the two steps are regarded as nucleotide pre-insertion and insertion [33, 34], respectively. In particular, the nucleotide insertion likely happens slowly [8, 9, 31], so we take it a rate-limiting step by default. Variation of the rate-limiting step in the scheme will be addressed later. Upon the nucleotide insertion, Watson-crick base pairing can form between the right nucleotide and the template.

Besides, in this five-state scheme, the catalytic part proceeds in two steps: The covalent linkage of the nucleotide ($IV \to V$) and PPi release ($V \to I$). Since the nucleotide selection happens before the end of the catalysis, splitting the PPi release from the catalysis or not does not matter to the selection. Essentially, one can identify four elementary selections along the reaction path (shown schematically in **Fig 4b**) prior to the product formation ($V$).

The first selection strategy, denoted $S_{III}^-$, rejects wrong nucleotides immediately upon binding. The selection strength can be written as $\eta_{III}^- = k_{III-}^w / k_{III-}^r$. The next selection strategy, denoted $S_{III}^+$, inhibits wrong nucleotides from *inserting* into the active site, at the strength of $\eta_{III}^+ = k_{III+}^r / k_{III+}^w$. The third selection strategy $S_{IV}^-$, at the strength of $\eta_{IV}^- = k_{IV-}^w / k_{IV-}^r$, destabilizes the wrong nucleotides after being inserted. The last selection strategy $S_{IV}^+$, at the strength of $\eta_{IV}^+ = k_{IV+}^r / k_{IV+}^w$, inhibits catalytic reaction of the wrong nucleotides. Each selection ends up with a reset of free energy difference between the wrong and right to $\delta_G$, as that in the enzyme-free case (see **Fig 4b**).

## IV.1 Error and speed control by the selection

To compare the speed and error rates under the above selection strategies, each selection is assumed to work at the same strength $\eta$, or use the same amount of differential free energy $\Delta = k_B T \ln \eta$. The polymerization rate $J/J_0$ vs. $\Delta$, the error rate *Err*, and the input-error ratio $1-i_r$ are plotted in **Fig 4 c-e**. Similar to the three-state scheme, one can see that the selection $S_{III}^-$ maintains the highest polymerization rate at a high NTP concentration (*e.g.* $J/J_0 \sim 0.8$). The polymerization rates under the selection $S_{III}^+$, $S_{IV}^-$ and $S_{IV}^+$ all approach to similarly low values when the selections get strong (*e.g.* $J/J_0 \sim 0.2$), alike that under $S_{III}^+$ in the three-state scheme.

One can also write down the output error rate, for simplicity, at the irreversible product release condition ($k_{I-} \to 0$)

$$Err = \frac{(1-i_r)[k_{III-}k_{IV-}k_{V+} + k_{III-}k_{IV-}k_{V+} + k_{III-}k_{IV+}k_{V+} + k_{III+}k_{IV+}k_{V+}]}{k_{III-}k_{IV-}k_{V-}(1-i_r+i_r\eta_G) + k_{V+}[k_{III-}k_{IV-} + k_{III-}k_{IV+} + k_{III+}k_{IV+} + i_r k_{III-}(k_{IV-}\eta_{III}^-\eta_{IV}^-\eta_{IV}^+ + k_{IV+}\eta_{III}^-\eta_{III}^+ - k_{IV-} - k_{IV+})]}$$ (4)

$\eta_G = e^{\delta_G}$ is to reset $\delta_G$ prior to the product formation; $\eta_G = 10$ is used by default.

One sees from **Eq 4** that the first two elementary selections $S_{III}^-$ and $S_{III}^+$ reduce the error rate through $\eta_{III}^- \eta_{III}^+$ such that the same error rates are obtained (under $\eta_{III}^- = \eta$ & $\eta_{III}^+ = 1$ and $\eta_{III}^- = 1$ & $\eta_{III}^+ = \eta$). Similarly, the latter two selections $S_{IV}^-$ and $S_{IV}^+$ also perform equally well in the error reduction. However, the performance of the latter two is inferior to the former two. Indeed, when only the first two selections work and are equally strong ($\eta_{III}^- = \eta_{III}^+ \gg 1$ and $\eta_{IV}^- = \eta_{IV}^+ = 1$),

$$Err = \frac{1-i_r}{i_r}[1 + \frac{k_{III-}k_{IV-}k_{IV+} + k_{III-}k_{IV+}k_{V+}}{k_{III-}(k_{IV-}+k_{IV+})k_{V+}}]\frac{1}{\eta_{III}^- \eta_{III}^+} = \frac{1-i_r}{i_r}[1+\kappa_{III}]e^{-2\Delta/k_B T}$$ (4a)

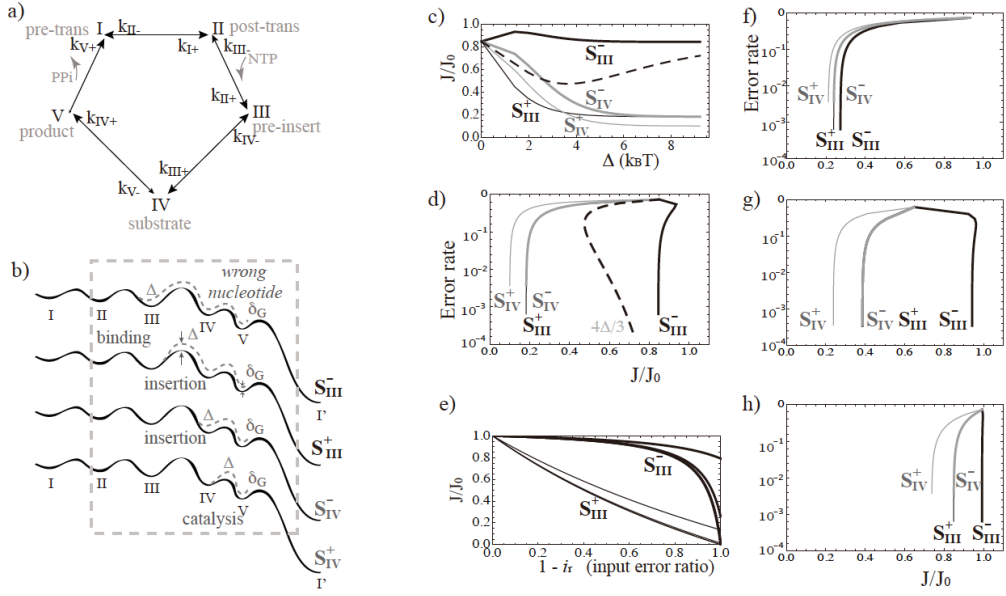

FIG. 4. Nucleotide selections in the five-state elongation scheme. (a) The five-state scheme consists of translocation, nucleotide pre-insertion (binding), insertion, catalysis, and PPi release. (b) Four elementary selections demonstrated on the free energy profile (for right / wrong nucleotides in solid / dashed line). $S_{III}^-$ enhances immediate rejection of the wrong nucleotides. $S_{III}^+$ inhibits the insertion of the wrong nucleotides. $S_{IV}^-$ rejects the wrong nucleotides upon base pairing at insertion, while $S_{IV}^+$ inhibits the catalysis of the wrong. $\Delta G_c^r - \Delta G_c^w = \delta_G$ is set at the end. (c) Polymerization rate vs. the differentiation free energy for selection $S_{III}^-$ (dark line), $S_{III}^+$ (dark thin line), $S_{IV}^-$ (gray line) and $S_{IV}^+$ (gray thin line). A combined selection strategy with all four elementary selections at equal strength $\eta = e^{\Delta/3k_BT}$ is also shown (dashed line). (d) The error rate vs. elongation rate under $S_{III}^-$, $S_{III}^+$, $S_{IV}^-$, $S_{IV}^+$, and the combined. (e) The polymerization rate vs. input error ratio $1-i_r$. For selections $S_{III}^-$ and $S_{III}^+$, lines at different selection strengths are shown ($\eta$ = 10, 10$^2$, 10$^3$, and 10$^4$ from up to down). For clarity, the lines for $S_{IV}^-$ and $S_{IV}^+$ are shown in Fig S4c in SM Appendix IV. (f-h) The error rate vs. elongation rate as in (d), while the rate-limiting step varies: (f), NTP concentration is low so that NTP binding is rate limiting; (g) The catalytic rate is low, and is much lower than the reverse rate of the nucleotide insertion; (h) The translocation is rate limiting.

In contrast, when only the two latter selections work and are equally strong ($\eta_{III}^- = \eta_{III}^+ = 1$ and $\eta_{IV}^- = \eta_{IV}^+ \gg 1$),

$$Err = \frac{1-i_r}{i_r}[1+\frac{k_{III-}k_{IV-}k_{IV+} + (k_{III-}+k_{III+})k_{V+}k_{V+}}{k_{III-}k_{IV-}k_{V+}}]\frac{1}{\eta_{III}^-\eta_{III}^+} = \frac{1-i_r}{i_r}[1+\kappa_{IV}]e^{-2\Delta/k_BT}$$

(4b)

Since $\kappa_{III} < \kappa_{IV}$, it leads to a smaller error rate in the first two selections, starting from state *III*, than that in the two selections from state *IV*. The feature is due to the linear reaction topology and is independent of the kinetic parameters (in addition, see **SM Appendix III** for a six-state scheme). When the last transition or the full scheme is reversible ($k_{I-} > 0$), $S_{III}^-$ and $S_{III}^+$ still perform equally well in error control, and outperform $S_{IV}^-$ and $S_{IV}^+$. For example, in **Fig 4d**, we see that the lowest error rate reached by $S_{III}^-$ or $S_{III}^+$ is ~6x10$^{-4}$ for $\Delta \sim 9$ k$_B$T. The error rate reached by $S_{IV}^-$ or $S_{IV}^+$ is ~ 4x10$^{-3}$ for the same $\Delta$. If one combines four elementary selections together, with each selection using a small differentiation free energy $\Delta/3=3$ k$_B$T (see **Fig 4d**), one also obtains a low error rate ~ 2x10$^{-4}$, while the speed is kept fairly high as $J/J_0 \sim 0.75$.

Similar to **Fig 3e**, **Fig 4e** shows how the polymerization rate changes as the input error ratio increases in solution. The polymerization rate under $S_{III}^-$ appears insensitive to input errors up to a quite high error portion (e.g. 80%), while that under the other selection strategies are not. The results for $S_{IV}^-$ and $S_{IV}^+$ are shown for clarity in **Fig S4c** in **SM Appendix IV**.

### IV.2 Kinetic impact and variation of the rate-limiting step

From the derivation, one can heuristically write down the elongation error rate (with an irreversible last step in the elongation scheme) in general as:

$$Err = \frac{(1-i_r)(\kappa_a + \sum_{M=m}^{N} a_M)}{(1-i_r)(\kappa_a + \sum_{M=m}^{N} a_M) + i_r(\kappa_b + \sum_{M=m}^{N} a_M \prod_{i=m}^{M} \eta_i^- \eta_i^+)}$$

(4c)

where $\kappa_a$, $\kappa_b$ and $a_M$ are combinations of the kinetic parameters. In current three-state scheme, $m = N = III$; in current five-state scheme, $m=III$, $N=IV$). Hence, as every elementary selection grows strong ($\eta_i^\pm = e^{\Delta_i^\pm/k_BT} \gg 1$), $Err \sim \frac{1-i_r}{i_r}(1+\kappa)e^{-\sum \Delta_i^{\ddagger}/k_B T}$ always holds. It clearly show that the error rate deceases exponentially with the accumulate differentiation free energy along the reaction path $\sum \Delta_i^{\ddagger}/k_B T$. If one lowers the rates of forward transitions involved in the selection (e.g. $k_{III+}$, $k_{IV+}$ or both), or raises the rates of backward transition, one can reduce the value of $\kappa$ and lower the error rate. Of course, this type of accuracy improvement is at price of lowering the speed, as that indicated in the efficiency-accuracy tradeoff.

One can also see how the speed and error rate control varies when the rate-limiting step varies in the elongation cycle. Here the rate-limiting step is determined according to the kinetics of the right substrate. First, if one lowers the NTP concentration, the nucleotide binding can become rate limiting. We see from **Fig 4f** that all selections significantly lower the speed below that without selection. The first selection $S_{III}^-$ also happens after the slow NTP binding, and cannot recover the speed to high. Next, if the catalysis becomes so slow such that the catalytic rate is much smaller than the reversal rate of the nucleotide insertion ($k_{IV+} \ll k_{IV-}$; see **Fig 4g**), then the error rate achieved by the former two selections ($S_{III}^-$ and $S_{III}^+$) become almost identical to that from the latter two ($S_{IV}^-$ and $S_{IV}^+$). Additionally, if the translocation after the product release happens quite slowly, then the elongation rate is dominated by the translocation rate, and cannot be reduced much by any nucleotide selection (see **Fig 4h**). Hence, the error and speed control patterns persist but become more or less pronounced at different rate-limiting conditions.

## V. DISCUSSION

In current work, we show how stepwise nucleotide selection could proceed efficiently for fidelity control in polymerase elongation. Basically, we want to identify selection strategies that achieve comparatively low error rates for certain free energy differentiation, without significantly lowering the polymerization speed. From previous studies, various ways of nucleotide selection had been reported for different polymerases [12]. As rate-limiting steps also vary among different systems, it is hard to identify common selection mechanisms. In this study, we demonstrate that for efficient selections, there exist some general features in the selection systems, as summarized below. However, we want to make it clear that polymerases are *not* necessarily evolved to be highly efficient in the selectivity. Their functional development has to meet various internal and external requirements.

To characterize stepwise nucleotide selection, one needs to consider the free energy differentiation between the right and wrong substrates at every checkpoint along the reaction path. The free energy differentiation $\Delta$ (>0) at any particular checkpoint relies on physical properties of the enzyme and ambient conditions. For example, to differentiate the substrate species, some structural or electrostatic characters of the protein have to be developed, while water molecules need to be more or less excluded, and certain ions may also be required for coordination [11, 35, 36]. Accordingly, the selectivity demands very specific and fine-tuning of molecular interactions, and the differentiation capacity is restricted at any one checkpoint ($\Delta$ cannot be very large).

Indeed, an elementary selection is either to inhibit the forward transition or to enhance the backward transition in the wrong substrate incorporation, by modulating the transition activation barrier by $\Delta$ in comparison to that of the right species. When $\Delta$ is fixed at every selection point, while the system kinetics varies in controlled conditions, the selection accuracy can be improved at compensation of the reaction efficiency.

On the other hand, if the reaction kinetics of the right substrates is given, while $\Delta$ is allowed to increase at one checkpoint, the error rate can be continuously lowered while the speed converges to a constant value. Depending on which selection checkpoint is exploited on the reaction path, the error rate and speed vary for a same value of $\Delta$. Our study shows that early selections on the reaction path outperform the late ones on the error reduction, and the initial selection is indispensible for maintaining the speed high. Here, the early selection starts upon the substrate binding, and the late one ends once the catalysis finishes. We essentially show that the error rate can be repeatedly lowered through the stepwise selection. That is to say, multiple kinetic checkpoints along the reaction path do improve the fidelity level as the free energy for differentiation accumulates. Mathematically, this property is similar to the amplifying effects in kinetic proofreading, as the elongation cycle is essentially maintained at the NESS with the detailed balance broken.

In previous sections, we compared error reduction and speed modulation of the elementary nucleotide selections in the three- and five-step elongation scheme. The three-state scheme is characterized by one-step NTP binding and two potential kinetic checkpoints, while two-step NTP insertion and four potential checkpoints apply in the five-state scheme. For mathematical simplicity, we assumed in both schemes that the PPi dissociation step is irreversible, as if PPi concentration is quite low around the active site of the polymerase. Similar results show as well in the fully reversible scheme on the speed and error control (see **SM Appendix II**). We also checked four-state and six-state kinetic schemes (see **SM Appendix III**). The four-state scheme is similar to the three-state scheme if there is only one-step NTP binding prior to the chemical catalysis; the scheme becomes similar to the five-state case if the NTP insertion happens in two steps. In the six-state scheme, we made a three-step pre-chemical NTP insertion process, with six potential checkpoints in total. The variations of the kinetic scheme lead to no essential changes in the speed and error control in the corresponding elementary selections.

Below, we summarize crucial aspects of the efficient fidelity control, which is to achieve low error rate without lowering much the speed. In the end, we apply current framework to describe some particular polymerases, and use this framework as well to analyze the information acquisition features of the selection system.

### V.1 Achieve low error rates – select early, properly, and repeatedly

We have examined elementary selections that tune only one transition barrier forward or backward at a time when incorporating the wrong nucleotides. The elementary selections are arranged sequentially along the reaction path of the elongation cycle. Our results highlight two interesting findings (*i*) The error rate achieved by the selection that rejects the wrong nucleotides from state *i* to *i*-1 ($S_i^-$ at the strength $\eta_i^- = e^{\Delta/k_BT}$), is always the same as that achieved by the selection right after, which inhibits the wrong nucleotides from transiting from state *i* to *i*+1 ($S_i^+$ at the same strength $\eta_i^+ = e^{\Delta/k_BT}$). This means $S_{III}^-$ and $S_{III}^+$, as well as $S_{IV}^-$ and $S_{IV}^+$ in the five-state scheme, perform equally well in the error reduction. It reveals that from a same kinetic or conformational state, one can conduct two equivalent strategies in fidelity control: The first strategy can be achieved usually by destabilizing the wrong substrate state configuration so that its backward transition becomes more likely; the second strategy relies on designing a harder barrier on the transition path for the wrong substrate state to move toward the next conformational state. (*ii*) The error rate achieved by the selection that inhibits the forward transition of the wrong nucleotides from *i* to *i*+1 ($S_i^+$ of $\eta_i^+ = e^{\Delta/k_BT}$) is always lower than that achieved by the selection that rejects the wrong substrates from next state *i*+1 to *i* back ($S_{i+1}^-$ of $\eta_{i+1}^- = e^{\Delta/k_BT}$). However, when the rate starting from *i*+1 forward is much smaller than that starting from *i*+1 back, the drop of the error reduction performance from $S_i^{\pm}$ to $S_{i+1}^{\pm}$ becomes insignificant. That is, when $k_{IV+} \ll k_{IV-}$ in the five-state scheme, the error rates becomes almost the same for all four elementary selections. In brief, the error reduction performance of the selection does not improve following down the reaction path. The property would persist in general to any elongation scheme.

As any nucleotide selection strategy can be regarded as a combination of the elementary selections, the above results give some rules of thumb on identifying a *proper* selection strategy. First, as a direct consequence of (*i*) above, one cannot combine a selection $S_i^-$ of strength $\eta_i^- = e^{\Delta/k_BT}$ with an 'anti-selection' $(S_i^+)^{-1}$ of strength $(\eta_i^+)^{-1} = e^{-\Delta/k_BT}$ (<1) to achieve an error reduction. Here the *anti-selection* indicates an operation with a 'selection' strength less than one, which favors the wrong substrate rather than the right one. The futile strategy is illustrated in **Fig 5a** *left*, as if the state *i* is destabilized for the wrong species, while *both* forward and backward barriers are reduced by Δ (>0). The operations $S_i^-$ and $(S_i^+)^{-1}$ simply cancel each other.

On the other hand, another selection strategy illustrated in **Fig 5a** *middle*, for example, does work well for an error reduction. It can be regarded as a combination of selection $S_i^+$ of strength $\eta_i^+ = e^{\Delta/k_BT}$ and an *anti-selection* $(S_{i+1}^-)^{-1}$ of strength $(\eta_{i+1}^-)^{-1} = e^{-\Delta/k_BT}$. $S_i^+$ outperforms $S_{i+1}^-$ in the error control as from (*ii*) above, so they do not fully cancel. They work together for the error reduction, though the performance is a bit inferior to $S_i^+$ alone, as $(S_{i+1}^-)^{-1}$ assists not the right but wrong nucleotide selection.

Consequently, the strategy shown in **Fig 5a** *right* works as a combination of $S_{i-1}^+$ of strength $\eta_{i-1}^+ = e^{\Delta/k_BT}$, $S_i^-$ of strength $\eta_i^- = e^{(\Delta'-\Delta)/k_BT}$ ($\eta_i^- > 1$ as $\Delta' > \Delta$), $(S_i^+)^{-1}$ of strength $(\eta_i^+)^{-1} = e^{(\Delta-\Delta')/k_BT}$ (<1), and $(S_{i+1}^-)^{-1}$ of strength $(\eta_{i+1}^-)^{-1} = e^{(\Delta-\Delta')/k_BT}$. Since $S_i^-$ and $(S_i^+)^{-1}$ cancel each other, they do not really work for selection. The overall strategy actually works as a combination of $S_{i-1}^+$ and $(S_{i+1}^-)^{-1}$, it reduces the error rate but the performance is a bit inferior to $S_{i-1}^+$ alone. Hence, the error rate achieved under this strategy cannot be lower than that achieved under a single elementary selection $S_{i-1}^+$ ($\sim e^{-\Delta/k_BT}$). It shows that an 'improperly' combined selection strategy cannot improve the fidelity over that of an elementary selection. It also indicates that to make a proper selection strategy to take advantage of every kinetic checkpoint, the separation between the free energy files of the wrong and right has to gradually expand along the reaction path.

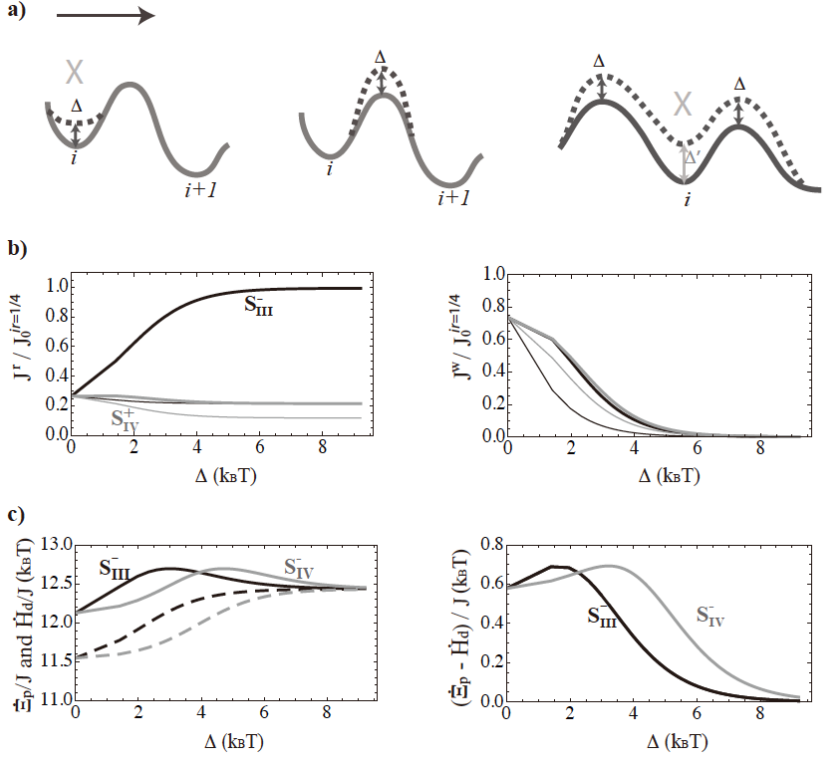

FIG. 5. Error reduction strategies, flux properties, and entropy changes. **(a)** Examples on futile and effective selection strategies. The one on the left is a futile strategy, which combines $S_i^-$ and $(S_i^+)^{-1}$ that cancel each other for error reduction. The one in the middle combines $S_i^+$ and $(S_{i+1}^-)^{-1}$ together and works for error reduction. The one on the right is a combination of $S_{i-1}^+$, $S_i^-$, $(S_i^+)^{-1}$, and $(S_{i+1}^-)^{-1}$; since $S_i^-$ and $(S_i^+)^{-1}$ cancel each other, the overall strategy is equivalent to $S_{i-1}^+$ and $(S_{i+1}^-)^{-1}$ together for error reduction. Note that $S^{-1}$ indeed selects *against* the right nucleotides instead of the wrong ones. **(b)** The elongation fluxes or rates for the right (*left*) and wrong (*right*) nucleotide species vs. the differentiation free energy $\Delta$, calculated from the five-state scheme. The normalization factor $J_0^{1/4}$ is the flux in the absence of the nucleotide selection at an input error ratio $1-i_r=3/4$ ($i_r=1/4$). **(c)** The entropy production per nucleotide incorporated ($\dot{\Xi}_P/J$, solid lines) and the heat dissipation per nucleotide ($\dot{H}_d/J$, dashed lines) vs. $\Delta$ (left), and the overall entropy change per nucleotide cycle ($\dot{\Xi}_P - \dot{H}_d$)/J vs. $\Delta$ (right).

Indeed, we obtained a general expression of the elongation error rate as a function of individual selection strength in **Eq 4c**. The properties (*i*) and (*ii*) summarized above are the natural consequences of this expression (due to the term $\sum_{M=m}^{N} a_M \prod_{i=m}^{M} \eta_i^- \eta_i^+$). Consequently, one obtains $Err \sim \frac{1-i_r}{i_r}(1+\kappa)e^{-\sum \Delta_i^{\pm}/k_B T}$ for any combined selection at the strong selection limit ($e^{\Delta_i^{\pm}/k_B T} \gg 1$). Importantly, it indicates that the error reduction can be amplified through multiple steps along the reaction path. Since $\kappa > 0$, it is easy to see that the error rate cannot be lower than $\frac{1-i_r}{i_r}e^{-\sum \Delta_i^{\pm}/k_B T}$, or the accuracy $A$ cannot be higher than $e^{\sum \Delta_i^{\pm}/k_B T}$. Lowering the forward rate or raising the backward rate for the involved transition in the selection can lower the value of $\kappa$, such that the error rate is reduced while the reaction slows down. When every elementary selection participates in the combined selection, $e^{\sum \Delta_i^{\pm}/k_B T}$ becomes the maximum accuracy as suggested previously [17]. Note that the maximum free energy differentiation $\sum \Delta_i^{\pm} = \Omega_{max}$ over the reaction path is well defined as long as $\Delta_i^{\pm} > 0$. When there exist $\Delta_i^{\pm} < 0$ at some checkpoints, however, $\sum \Delta_i^{\pm} = \Omega_{max}$ does not coincide with the seemingly largest free energy difference between the right and wrong (e.g. $\Delta'$ is the largest difference in **Fig 5a** right, but is indeed irrelevant to the selectivity). Hence, an overall description of the stepwise selection using only the maximum free energy differentiation can be insufficient or even misleading.

### V.2 Maintain high polymerization speed – initial screening is indispensible

Our results indicate that the overall elongation rate is more or less reduced upon any nucleotide selection. It is consistent with the understanding that raising the accuracy is achieved at the price of lowering the speed. Indeed, the sort of tradeoff idea applies for a certain differentiation capacity. When the energy differentiation increases for any selection checkpoint, the error rate can be continuously lowered while the speed converges to a certain value. Nevertheless, for a constant energy differentiation capacity, varying the selection checkpoint can possibly improve both the fidelity and speed. For the very first selection, it outperforms the later selections not only in achieving the low error rate, but also in maintaining high speed.

**Fig 5b** shows the respective polymerization rates of right and wrong nucleotides, as $\Delta$ increases for various elementary selections (as in the five-state scheme). When there is no selection ($\Delta$=0), the relative flux for the right species ($J^r/J_0^{1/4}$) is 0.25, while that for the wrong species ($J^w/J_0^{1/4}$) is 0.75, as the input error ratio is ¾ ($i_r$=1/4). When $\Delta$ increases, the wrong fluxes uniformly decrease to zero. On the other hand, the right fluxes diminish to small values for all but the initial screening ($S_{III}^-$). The more stringent the initial screening, the higher the polymerization flux of the right nucleotides, as most of the wrong species are expelled away soon at entry and replaced by the right species. However, the initial free energy differentiation can be quite limited: When the nucleotide is just recruited, the site is relatively open and water solvent is not well excluded yet. Hence, it is unlikely to achieve nucleotide selection largely at the beginning. Nevertheless, when initial screening is *combined* with selections performed later in the cycle, the polymerization speed would still be maintained high, approaching to the speed under the initial screening alone. This is because the presence of the initial screening makes it efficient to throw away the wrong substrates in the proofreading-free system. Hence, even a small portion of initial screening in the combined selection can lead to fairly robust polymerization rates insensitive to input error rates from the solution. The properties highlight the importance of *including* the substrate screening at the beginning for an efficient selection, even there is only a limited amount of free energy differentiation.

When the nucleotide concentration is very low, however, even the initial selection can lower the elongation rate significantly. This is because the selection happens after the rate-limiting nucleotide binding, and the replacement of the right nucleotides slows down. On the other hand, if the rate-limiting step happens far behind all selection checkpoints, such as at the translocation, then the selection hardly impacts on the speed. Anyhow, it has not been reported the translocation being the single rate-limiting event, though it can be similarly slow as some catalytic or pre-catalytic event [37]. In brief, for polymerase elongation at high nucleotide concentration, the initial selection always helps to maintain the speed high, while later selections lower the speed. This gives some clues to determine essential amino acids for the initial screening through mutagenesis: For mutations that adversely affect the accuracy and speed, the original amino acids at the mutation sites likely contribute to the initial screening; for those that lower the accuracy but not the speed, either the original amino acids do not contribute to the initial selections, or the mutations still keep the initial screening on.

### *V.3 The nucleotide selection in some exemplary polymerase systems*

In this work, we have used elongation kinetic data of single-subunit T7 RNAP [4, 7, 31] for numerical demonstration (see kinetic parameters in **Table S1** from **SM Appendix IV**). Though the three-state scheme had been employed in early work for experimental data fitting [4], later on studies supported a five-state elongation scheme in this system [31]. In the five-state scheme, the pre-chemical nucleotide insertion following the initial nucleotide binding/pre-insertion is regarded rate limiting [31]. Our recent molecular dynamics simulations show that substantial nucleotide selection happens prior to the full insertion of the nucleotide into the active site in T7 RNAP [38]. That is, both the initial screening $S_{III}^-$ upon NTP per-insertion and the second selection $S_{III}^+$ during the NTP insertion play essentials roles in the nucleotide selection. Hence, T7 RNAP seems to be a quite efficient selection system that can fully employ the early selections on the reaction path. Since the error rate achieved by T7 RNAP is ~ $10^{-4}$ [32], one can estimate the maximum or accumulate free energy differentiation at ~ 10 $k_B T$. Actually, T7 RNAP achieves the error rate without proofreading detected. This likely explains why the nucleotide selection has to be efficient in this RNAP.

Next, we examined selection kinetics in T7 DNAP as the kinetic rates for incorporating both the right and wrong nucleotides had been reported in this system [14] (see **SM Appendix IV**). The chemical catalysis proceeds more slowly than the nucleotide insertion in T7 DNAP, while the reversal of the nucleotide insertion happens extremely slowly. The selection strengths are identified as: $\eta_{III}^-$ ~7 for $S_{III}^-$, $\eta_{III}^+$ ~ 3 for $S_{III}^+$, $\eta_{IV}^-$ ~263 for $S_{IV}^-$, and $\eta_{IV}^+$ ~ 1200 for $S_{IV}^+$, giving the differentiation free energies $\Delta_i$ /$k_B T$ as {1.9, 1.1, 5.6, 7.1}. The DNAP conducts proper nucleotide selection by combining all four elementary selections. However, the initial screening does not seem to be strong enough to support very high speed (see **SM Appendix IV**); the first two selections also do not appear strong enough to make the overall selection highly efficient. The full selection gives an error rate ~ $10^{-3}$ in T7 DNAP. The performances seem to leave room for further improvements by proofreading, which is indeed required and substantial in this DNAP.

In multi-subunit RNAPs, recent mutagenesis studies nicely show that discrimination against wrong nucleotides proceeds via a stepwise mechanism, and each step contributes differently to the overall fidelity [26]. In these systems, at least two steps happen prior to the chemical catalysis [39], starting from the NTP entry. In particular, the non-complementary NTPs are discriminated efficiently through both the first and second checkpoints, at the open active center and through a trigger loop folding process, respectively. The discrimination of the deoxy- NTPs does not happen until after the first checkpoint, hence, appearing less efficient. Indeed, the regulation of the enzyme activities is largely controlled through the trigger loop folding, providing possibilities that several selection checkpoints are coordinated in the fidelity control. It is not clear which step is rate limiting in the multi-subunit RNAPs. Likely one slow event takes place before or during catalysis, and another slow event is around translocation stage [37]. In that case, the speed modulation of the nucleotide selection may not be significant.

### V.4 Entropy production and heat dissipation under nucleotide selection

Last, we use current framework to quantify entropy production and energy dissipation under the nucleotide selection during the elongation NESS [40, 41]. These quantities are physically important to the selection system at nano-scale, but are not well defined in conventional simplified kinetic studies. Consider that the selection can never be perfect to prevent all errors, one expects at least two species (right and wrong) identified during the elongation. In solution of the mixed nucleotides, however, there is no way to identify the nucleotide species, so all input substrates are regarded as one species. Hence, during the template-based polymerization process, one always expects entropy variations upon the nucleotide species recognition.

The entropy variations can be decomposed into two components, the entropy production and heat dissipation [40, 42], with their respective rates denoted $\dot{\Xi}_p$ and $\dot{H}_d$. We derived both quantities in **Appendix V in SM** as:

$$\dot{\Xi}_p = J[\Delta G_c + (1-Err)k_BT\ln\frac{i_r}{1-Err} + Err\cdot(k_BT\ln\frac{1-i_r}{Err}-\delta_G)] \quad \textbf{(5a)}$$

$$\dot{H}_d = J\{\Delta G_c + (1-Err)k_BT\ln i_r + Err\cdot[k_BT\ln(1-i_r)-\delta_G]\} \quad \textbf{(5b)}$$

where $\Delta G_c \equiv \Delta G_c^0 + k_BT\ln\frac{[NTP]}{[PP_i]}$ is the overall free energy input. Hence, the net entropy change rate can be written down as $\dot{\Xi}_p - \dot{H}_d = Jk_BT[(1-Err)\ln\frac{1}{1-Err} + Err\ln\frac{1}{Err}]$.

Here the protein-solution contribution to the entropy production was not counted in. We focus only on the information content of the polymer chain. The chain disorder can indeed drive the polymer growth [28]. The entropy production and heat dissipation rates vs. the differentiation free energy (as in the five-state scheme) are provided in **Fig S5 in Appendix V**. In current example, the highest overall entropy change rate is close to ~80 $k_BT$/s at $\Delta \sim 2$ $k_BT$ under the initial screening $S_{III}^-$. The corresponding entropy production rate is close to 1500 $k_BT$/s, with most part being dissipated as heat.

The entropy production and heat dissipation per nucleotide cycle ($\dot{\Xi}_p/J$ or $\dot{H}_d/J$) are shown in **Fig 5c** left. The entropy production increases with the selection strength and then decreases. The heat dissipation always increases with the selection strength, until it converges to the entropy production. As $Err \to 0$, the net entropy change approaches zero (one *unidentifie*d species in, and one *right* species out).

In **Fig 5c** right, we show the net entropy change per nucleotide cycle $(\dot{\Xi}_p - \dot{H}_d)/J$. It is easy to see that the net entropy change is positive and bound by $k_BT\ln 2$ per nt as in any two-digit system (right and wrong). That says, the system entropy reaches to a maximum with equal portions of the nucleotide species incorporated. Beyond that, strong nucleotide selection quenches the net entropy production. Accordingly, upon week substrate selection (assuming the input nucleotides are dominated by the wrong species), the early selections ($S_{III}^-$ and $S_{III}^+$) on the reaction path promote entropy production more efficiently than the late selections ($S_{IV}^-$ and $S_{IV}^+$) on the reaction path. While upon the strong nucleotide selection, the early selections become more efficient to quench the entropy production. On the other hand, the early selections always facilitate the heat exchanges with the environment, comparing with the late selections.

## VI. CONCLUSIONS

In this work we have studied how stepwise nucleotide selection could proceed efficiently during template-based polymerase elongation. Basically, the selection can happen at multiple checkpoints prior to the end of chemical catalysis or the product formation. At each state, conformational transition of the enzyme backward or forward is enhanced or inhibited when the enzyme is bound with a wrong nucleotide. The selection through a single backward or forward transition is regarded as an elementary selection, and any selection in general can be regarded as a combination of the elementary selections. An efficient selection strategy takes advantage of multiple selection checkpoints to reduce the error rate repeatedly, and selects as early along the reaction path. At the same time, any selection checkpoint is subject to structural and energetic constraints to differentiate the nucleotide species. The efficient selection strategy achieves a low error rate with a limited amount of differentiation free energy accumulated along the reaction path, while minimally perturbs the overall elongation rate or speed.

We found that at the sufficiently high nucleotide concentration, the initial screening selecting against wrong nucleotides immediately upon their arrival perturbs the elongation rate slightly, while selections thereafter on the reaction path can significantly diminish the elongation rate.

Importantly, combing the initial screening with selections afterwards keeps the speed similarly high as that under the initial screening alone. Hence, for polymerases that need high cycling rates, the initial screening seems indispensible, and even a small differentiation there can help. Interestingly, we found that the early selections along the reaction path outperform the late ones in the error reduction, as lower error rates are achieved under the early rather than the late selections at the same free energy differentiation. In particular, for a pair of neighboring elementary selections, the one rejects the wrong substrate state back to the previous state and the one inhibits it toward the next state give a same error rate at the same free energy differentiation. These properties persist but become more or less pronounced at different rate-limiting conditions. In counting the entropy production rate of the selection system, we notice that a large portion of the entropy production is dissipated as heat in maintaining the elongation far from equilibrium. Comparing to the late selections, the early selections promote the information entropy production when the selection is weak, while quench the entropy production when the selection gets strong.

Based on this framework, we compared a proofreading-free T7 RNAP with a proofreading T7 DNAP. We found T7 RNAP to be an efficient selection system while T7 DNAP does not seem so. The current work of the stepwise nucleotide selection supports further quantitative researches to reveal underlying mechanisms of the selection. It may further help molecular engineering and redesign of efficient selection systems.

# ACKNOWLEDGEMENTS

Current work is supported by NSFC under the grant No. 11275022. Thanks Dr Hong Qian for helpful discussions. Thanks Dr George Oster and Dr Yuhai Tu for reading and commenting on the manuscript. Finally thanks the anonymous reviewers for constructive suggestions.

# Supplementary Material for "Efficient fidelity control by stepwise nucleotide selection in polymerase elongation"

## Appendix I: The generic three-state scheme and the efficiency-accuracy tradeoff

In Eq 2 in main text, the polymerization flux or rate $J$ is obtained in the Michaelis-Menten form. The maximum rate constant $k_{max}^0$ and the Michaelis constant $K_M^0$ are written as:

$$k_{max}^0 = \frac{k_{I+} k_{III+}}{k_{I+} + k_{III+}} \tag{S1}$$

$$K_M^0 = \frac{k_{I+} + k_{II-}}{k_{I+} + k_{III+}} \frac{k_{III+} + k_{III-}}{k_T^0} \tag{S2}$$

where $k_T^0$ is the NTP binding constant ($k_{II+} = k_T^0 [NTP]$). Correspondingly, the exact forms of $\Gamma$ and $\Lambda$ are:

$$\Gamma = \frac{(1-i_r)(\eta_{III}^+ - 1)(k_{III+} + k_{III-})}{k_{III+} + (1 - i_r \eta_{III}^- \eta_{III}^+) k_{III-}} \cdot \frac{k_{I+}}{k_{I+} + k_{III+}} \approx \frac{(1-i_r)(\eta_{III}^+ - 1)(k_{III+} + k_{III-})}{k_{III+} + (1 - i_r + i_r \eta_{III}^- \eta_{III}^+) k_{III-}} \tag{S3}$$

$$\Lambda = \frac{k_{III+} + k_{III-} \eta_{III}^- \eta_{III}^+}{k_{III+} + (1 - i_r + i_r \eta_{III}^- \eta_{III}^+) k_{III-}} \tag{S4}$$

The approximation in Eq S3 is taken for $k_{III+} \ll k_{I+}$, that is, the translocation rate $k_{I+}$ is much larger than the catalytic rate $k_{III+}$. It had been measured that the translocation is much faster than other kinetic steps in the elongation cycle [1,2]. If the translocation slows down, the value of $\Gamma$ will decrease, and the impact from $\Gamma$ weakens. As $k_{III+} \gg k_{I+}$, $\Gamma \to 0$, the selection strength $\eta$ can only modulate $K_M$ rather than $k_{max}$ in the Michaelis-Menten form of the elongation rate. In that case, the nucleotide insertion happens very fast and the selection cannot affect much the elongation rate at fairly high nucleotide concentration.

To make it clear how the individual selection strength $\eta = e^{\Delta/k_B T}$ or the differentiation free energy $\Delta$ affects $\Gamma$ and $\Lambda$ to modulate the rate and efficiency of the polymerization, **Fig S1** below shows $\Gamma$, $\Lambda$, $k_{max} = \frac{k_{max}^0}{1+\Gamma}$, $K_M = \frac{\Lambda}{1+\Gamma} K_M^0$, and $\zeta \equiv \frac{k_{max}}{K_M} = \frac{1}{\Lambda} \zeta_0$ vs. $\Delta$ under the two elementary selections $S_{III}^-$ and $S_{III}^+$ in the three-state scheme. Note that the two selections give the same curves of $\Lambda$ and $\zeta = k_{max}/K_M$. Indeed, $\Lambda, \zeta$, and the error rate $Err$ all depend only on the accumulate selection strength $\eta_{III}^- \eta_{III}^+$ rather than the individual terms.

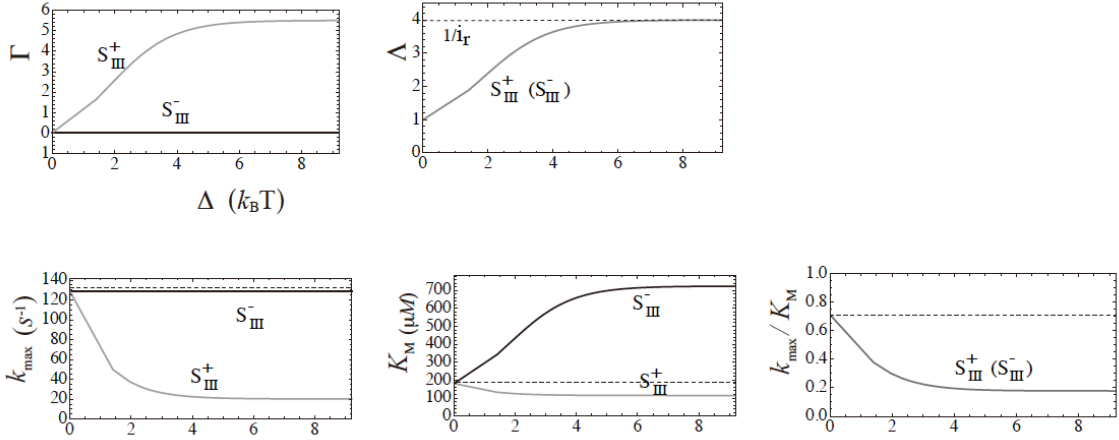

**Fig S1**. Variations of $\Gamma$, $\Lambda$, $k_{max}$, $K_M$, and $\zeta = k_{max}/K_M$ upon the variation of the differentiation free energy $\Delta$ under respective selection $S_{III}^-$ (black) and $S_{III}^+$ (gray) in the three-state elongation scheme.

As in the efficiency-accuracy tradeoff as discussed early [3-5], we can calculate straightforward the efficiency $\zeta^r$ for incorporating the right nucleotide substrates:

$$\zeta^r = \frac{k_{cat}^r}{K_M^r} = \frac{\frac{\Lambda}{1+\Gamma}k_{max}^0}{\frac{\Lambda}{1+\Gamma}K_M^0} = \zeta^0 = \kappa_a \cdot \frac{k_{III+}}{k_{III+}+k_{III-}} \tag{S5}$$

where $\kappa_a \equiv \dfrac{k_T^0 k_{I+}}{k_{I+}+k_{II-}} = \dfrac{k_T^0}{1+\dfrac{k_{II-}}{k_{I+}}}$ is the effective binding constant in the three-state cycle, $\kappa_a \sim k_T^0/2$ as polymerases translocate in a Brownian ratchet fashion, with equal forward and backward rates ($k_{I+} \sim k_{II-}$) [6]. From Eq 3a in main, we write

$$\frac{d-A}{d-1} = \frac{\eta_{III}^- \eta_{III}^+ - (k_{III-}\eta_{III}^-\eta_{III}^+ + k_{III+})/(k_{III-}+k_{III+})}{\eta_{III}^-\eta_{III}^+ - 1} = \frac{k_{III+}}{k_{III-}+k_{III+}},$$ 

where $d \equiv \eta_{III}^-\eta_{III}^+ = e^{\Omega_{max}/k_BT}$ is the accumulate selection strength, or the 'maximum accuracy'.

Hence, we obtain the linear efficiency-accuracy tradeoff relationship as that in Eq [1] from [7].

$$\zeta^r = \frac{k_{cat}^r}{K_M^r} = \kappa_a \frac{d-A}{d-1} \tag{S6}$$

## Appendix II: In the reversible three-state scheme of polymerase elongation

One can solve Eq 1 in main at the steady state $\frac{d}{dt}\Pi = 0$ *without* using the approximation $k_{l-} \to 0$. The error rate *Err* can be obtained iteratively. From the expression of *Err* (not shown as it is too long), one can see it contains *only* $\eta_{III}^{-}\eta_{III}^{+}$ but not the individual term $\eta_{III}^{-}$ or $\eta_{III}^{+}$. Hence, the same strength of the individual selection always leads to a same error rate. In **Fig S2** below, we show how the selections impact on the speeds and error rates:

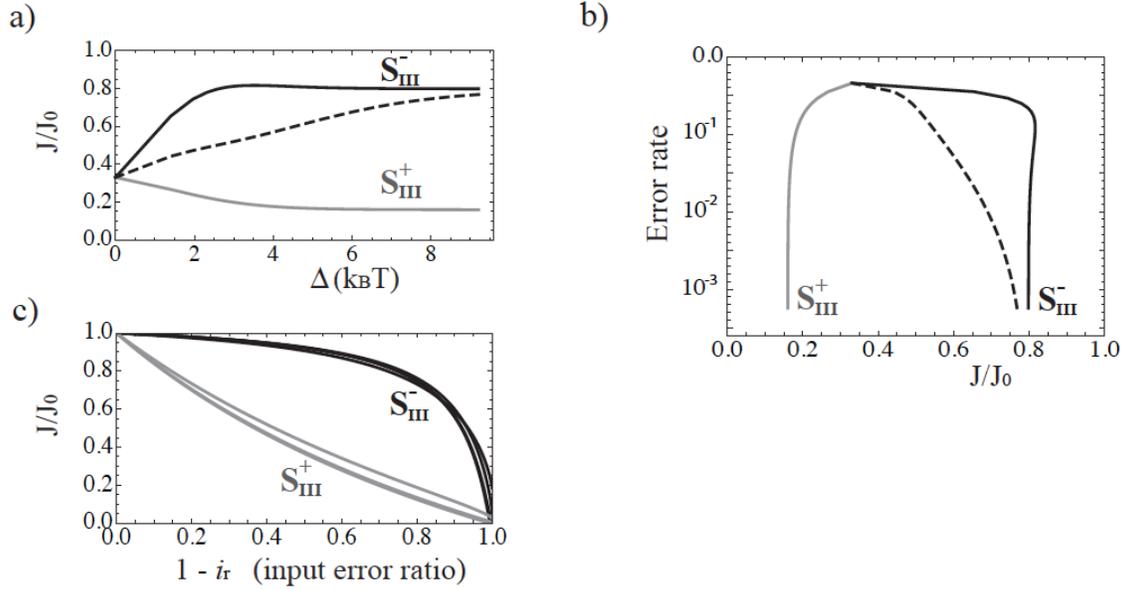

***Fig S2*** *The polymerization rates/speeds and error rates under the two elementary selections in the fully reversible three-state scheme. (a) The polymerization rate (normalized $J/J_0$) vs. the differentiation free energy $\Delta = k_B T \ln \eta$. (b) The error rate vs. the polymerization rate as $\Delta$ varies. (c) The polymerization rate vs. the input error ratio. See detailed captions in Fig 3 (c-e).*

From the results above, we see that the conclusions still hold as: (1) The initial selection $S_{III}^{-}$ leads to a polymerization rate close to $J_0$, much higher than that under $S_{III}^{+}$. Under the combined selection, the polymerization rate also approaches to that under $S_{III}^{-}$ alone as the selection gets strong. (2) $S_{III}^{-}$ and $S_{III}^{+}$ perform equally well in the error control, i.e., giving the same error rate at the same differentiation free energy or selection strength. (3) The polymerization rates are fairly insensitive to changes of the input error rate under selection $S_{III}^{-}$.

## Appendix III: A six-state scheme with three pre-chemistry transitions

One can also build a six-state kinetic scheme by putting another 'insertion' step (see **Fig S3a** below) prior to the chemical catalysis, in addition to the pre-insertion and insertion step in the five-state scheme. Solving an equation similar to Eq 1 but in a ten-state vector space: $\Pi =(P_I, P_{II}, P_{III}^r, P_{III}^w, P_{IV}^r, P_{IV}^w, P_V^r, P_V^w, P_{VI}^r, P_{VI}^w)^T$ ('r' and 'w' labeling for probabilities of the wrong and right nucleotide bound states, respectively), one obtains the error rate:

$$Err = \frac{1}{1+\dfrac{i_r[k_{VI+}(k_{III+}k_{IV+}k_{V+}+k_{III-}\eta_{III}^-\eta_{III}^+(k_{IV+}k_{V+}+k_{IV-}\eta_{IV}^-\eta_{IV}^+(k_{V+}+k_{V-}\eta_V^-\eta_V^+)))+k_{III-}k_{IV-}k_{VI-}\eta_G]}{(1-i_r)[k_{III-}k_{IV-}k_{V-}k_{VI-}+k_{VI+}(k_{III-}k_{IV+}k_{V+}+k_{III-}(k_{IV+}k_{V+}+k_{IV-}k_{V-}+k_{IV-}k_{V+}))]}} \quad (S7)$$

The diagrams on polymerization rates and error rates under the selections are:

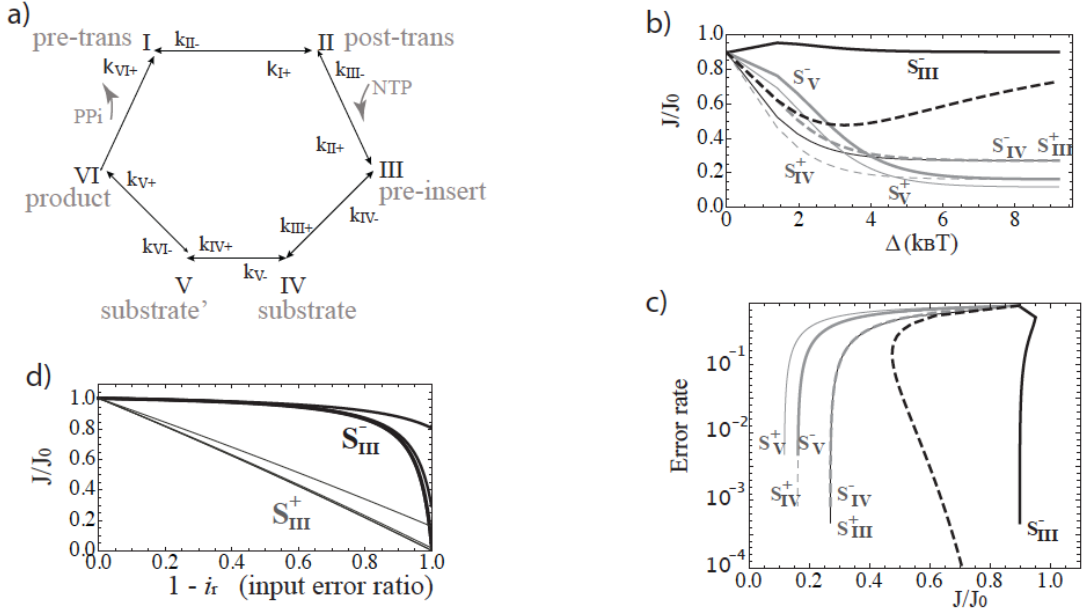

**Fig S3**. *The polymerization rates and error rates under nucleotide selections in the six-state scheme. (a) The six-state elongation scheme with three pre-chemistry steps: pre-insertion (II→III), two insertions (III→IV and IV→V) (b) The polymerization rate (normalized J/J₀) vs. $\Delta = k_B T \ln \eta$. (c) The error rate vs. the polymerization rate as $\Delta$ varies. (d) The polymerization rate vs. the input error ratio. See detailed captions in Fig 4 (c-e).*

From the above results we see that the conclusions listed in the previous section still hold. In particular, one sees that two neighboring selections against wrong nucleotides, $S_i^-$ and $S_i^+$ from the same initial state **i**, give the same error rates at the same selection strength (as $\eta_i^- = \eta_i^+$).

## Appendix IV: T7 DNAP in five-state scheme

For comparison, kinetic data utilized for T7 RNAP [8, 9] are listed in **Table S1** below; kinetic data for T7 DNAP [10] are listed in **Table S2**, for incorporating both right and wrong nucleotides.

| Translocation | | NTP binding/ pre-insertion | | NTP insertion | | Chemical catalysis | | PPi dissociation | |
|---|---|---|---|---|---|---|---|---|---|
| $k_{t+}$ | $k_{t-}$ | $k_{b+} = k_o^T[NTP]$ | $k_{b-}$ | $k_{i+}$ | $k_{i-}$ | $k_{c+}$ | $k_{c-}$ | $k_{d+}$ | $k_{d-}$ |
| 5000 | 5000 | 2*588 | 2x80=160 | **220** | 210 | 1000 | 135 | **1200** | 0.01 |

**Table S1** Kinetic rates for T7 RNAP used in current work for numerical demonstration by default. All rates listed above are in the unit of s[-1]. Data in **bold** are from transient state kinetics measured[8]. Other were numerically tuned or used for convenience[9].

| Translocation | | NTP binding/ pre-insertion | | NTP insertion | | Chemical catalysis | | PPi dissociation | |
|---|---|---|---|---|---|---|---|---|---|
| $k_{t+}$ | $k_{t-}$ | $k_{b+} = k_o^T[NTP]$ | $k_{b-}$ | $k_{i+}$ | $k_{i-}$ | $k_{c+}$ | $k_{c-}$ | $k_{d+}$ | $k_{d-}$ |
| 5000 | 5000 | 2*588 | 2x **28**=56 | **660** | **1.6** | **360** | 320 | 1200 | 0.01 |
| 5000 | 5000 | 2*588 | 2x**200**=400 | **220** | **420** | **0.3** | 320 | 1200 | 0.01 |

**Table S2** Kinetic rates for T7 DNAP used for numerical demonstration below (upper row for the right substrate and lower row for the wrong substrate). All rates listed above are in the unit of s[-1]. Data in **bold** are from reference [10]. Other data are set the same as that in T7 RNAP (see **Table S1**) for an easy comparison.

The selections in T7 DNAP, in comparison with that of T7 RNAP (see **Fig 4** and **5** in main), are shown below in **Fig S4**. For clarity, one diagram not shown in **Fig 4e** for T7 RNAP is now shown in **Fig S4c**, while the rest of all diagrams (a, b, d, and e) are for T7 DNAP. The combined selection strategy for T7 DNAP is shown in **Fig S4a** (dashed line). In this strategy, the strengths for three of the four elementary selections are set at $\eta_{III}^+ = 3$ $\eta_{IV}^- = 262.5$ $\eta_{IV}^+ = 1200$ for $S_{III}^+$, $S_{IV}^-$, $S_{IV}^+$, while the strength of the initial selection $S_{III}^-$ varies. The arrow points to $\Delta \sim 2$ $k_B$T for $S_{III}^-$ as reading the data from **Table S2**, giving a polymerization rate $\sim 48$ nt/s ($J_0 \sim 144$ nt/s). One sees that if $\Delta$ increases upon the initial screening $S_{III}^-$, the polymerization rate would increase. Overall, the nucleotide selection in T7 DNAP gives an error rate about 10[-3]. The proofreading is expected for further error reduction in T7 DNAP.

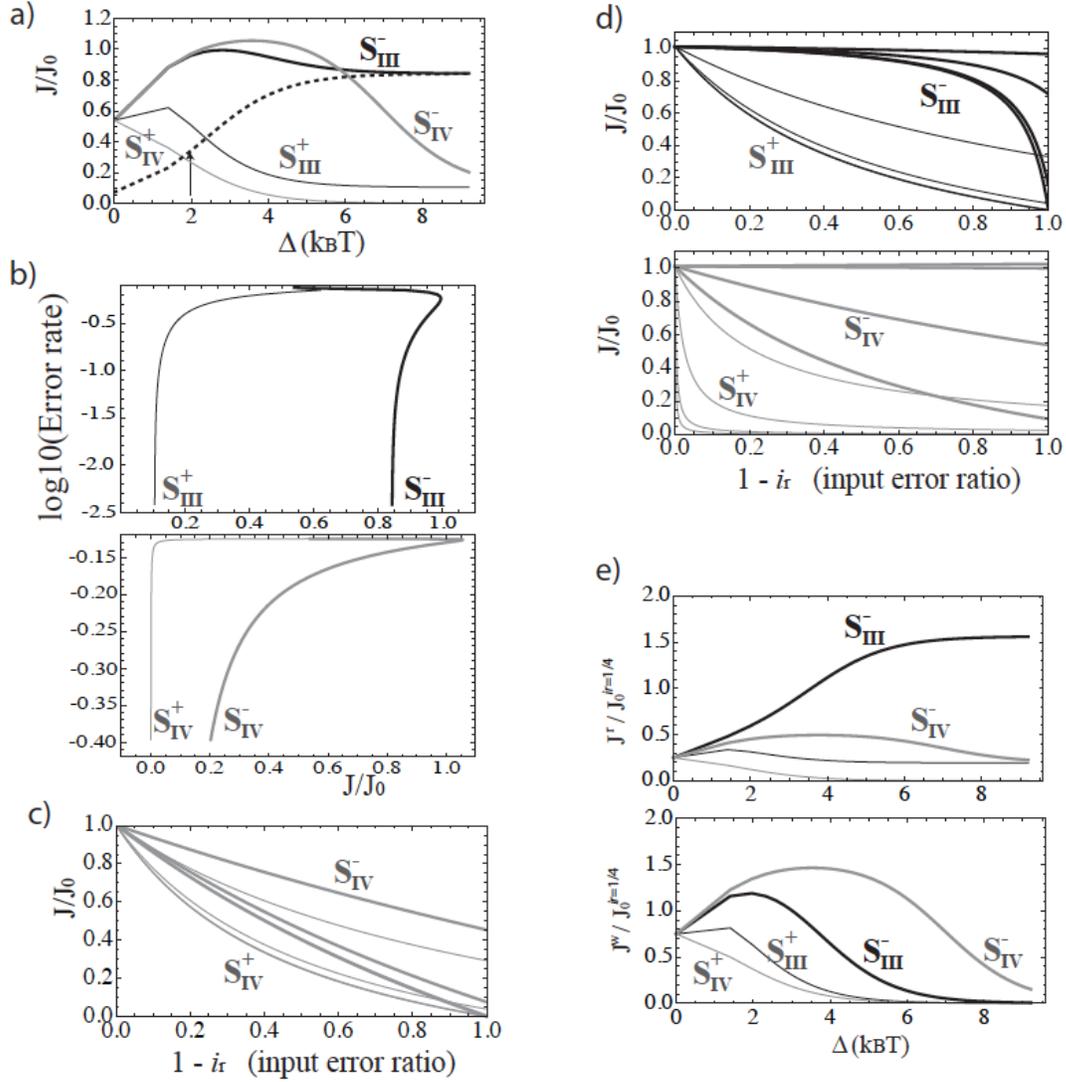

*Fig S4*. The polymerization rate and error rate under nucleotide selection in the five-state scheme of T7 DNAP. (a) The polymerization rate (normalized $J/J_0$) vs. $\Delta = k_B T \ln \eta$. The dashed line is the combined nucleotide selection, as $\eta_{III}^+ = 3$ $\eta_{IV}^- = 262.5$ $\eta_{IV}^+ = 1200$ while $\Delta$ varying for $S_{III}^-$. The arrow indicates $\Delta \sim 2k_B T$ ($\eta_{III}^- = 7.1$) for the initial screening selection. (b) The error rate vs. the polymerization rate as $\Delta$ varies. Shown in two diagrams for clarity. (c) The polymerization rate vs. the input error ratio under selection $S_{IV}^-$ *and* $S_{IV}^+$ *for T7 RNAP* as that was not shown in main **Fig 4e**. (d) The polymerization rate vs. the input error ratio under selections in T7 DNAP. One can compare the results (a, b, d) with that in main **Fig 4 (c-e)**. (e) The polymerization rates for the right and wrong nucleotide species in T7 DNAP, in comparison with that of T7 RNAP in main **Fig 5b**.

## Appendix V: Entropy production and heat dissipation

Below we discuss how nucleotide selections impact on the entropy production and heat dissipation, or the net entropy change. The entropy production rate at the NESS is counted as [11]

$$\dot{\Xi}_p = k_B T \sum_{i=1}^{N} (J_{i,i+1} - J_{i+1,i}) \ln(J_{i,i+1} / J_{i+1,i}) \quad (S8)$$

where $i$, $i+1$ represent two consecutive discrete states in the enzymatic cycle (with $N$ kinetic states, and $N+1$ reset to 1). In particular, the forward and backward probability fluxes are $J_{i,i+1} = P_i k_{i+}$ and $J_{i+1,i} = P_{i+1} k_{i+1-}$, while the net flux (i.e., the polymerization rate) is $J = J_{i,i+1} - J_{i+1,i}$. $J \equiv const > 0$ is maintained at the NESS.

When there is no error or wrong species, $\dot{\Xi}_p^0 = J_0 \Delta G_c$, since $\sum_{i=1}^{N} \ln(P_i / P_{i+1}) = 0$ ($P_{N+1} \equiv P_1$), where $\Delta G_c \equiv k_B T \sum_{i=1}^{N} \ln \frac{k_{i+}}{k_{i+1-}}$ is the chemical free energy input. At the same time, the heat dissipation rate is counted as

$$\dot{H}_d = J k_B T \sum_{i=1}^{N} \ln \frac{k_{i+}}{k_{i+1-}} \quad (S9)$$

so that $\dot{H}_d^0 = J_0 \Delta G_c = \dot{\Xi}_p^0$ in the absence of the error.

When there are both right and wrong nucleotides in competition, one can count $\dot{\Xi}_p$ for the right and wrong as
$$\dot{\Xi}_p^r = J^r (k_B T \ln \frac{i_r P_{II}}{P_{II}^r} + \Delta G_c) = J(1 - Err)[\Delta G_c^r - k_B T \ln(1 - Err)] \quad \text{and} \quad \dot{\Xi}_p^w = J \cdot Err \cdot (\Delta G_c^w - k_B T \ln Err) ,$$

where $\Delta G_c^r \equiv \Delta G_c + k_B T \ln i_r$ and $\Delta G_c^w \equiv \Delta G_c + k_B T \ln(1 - i_r) - \delta_G$ are the free energy input for a right nucleotide and a wrong one. In particular, $\Delta G_c \equiv \Delta G_c^0 + k_B T \ln \frac{[NTP]}{[PPi]}$ includes a standard free energy input of the nucleotide incorporation $\Delta G_c^0$, and the overall NTP/PPi concentration dependent part. The total entropy production rate is then counted as $\dot{\Xi}_P = \dot{\Xi}_p^r + \dot{\Xi}_p^w$:

$$\dot{\Xi}_p = J[\Delta G_c + (1 - Err) k_B T \ln \frac{i_r}{1 - Err} + Err \cdot (k_B T \ln \frac{1 - i_r}{Err} - \delta_G)] \quad (S10)$$

When $Err \to 0$, $\dot{\Xi}_P$ converges to $J \Delta G_c^r$. Correspondingly, the heat dissipation rate now is $\dot{H}_d = J[(1 - Err) \Delta G_c^r + Err \cdot \Delta G_c^w]$:

$$\dot{H}_d = J\{\Delta G_c + (1 - Err) k_B T \ln i_r + Err \cdot [k_B T \ln(1 - i_r) - \delta_G]\} \quad (S11)$$

As a result the overall entropy change rate is

$$\dot{\Xi}_p - \dot{H}_d = Jk_BT[(1-Err)\ln\frac{1}{1-Err} + Err\ln\frac{1}{Err}] \qquad (S12)$$

with the first and second terms coming from the right and wrong species, respectively.

The respective changes of entropy production rate $\dot{\Xi}_P$ and heat dissipation rate $\dot{H}_d$ along with $\dot{\Xi}_P - \dot{H}_d$ can be found in **Fig S5** below, as the selection gets strong for each of the four elementary selections in the five-state kinetics. The changes are dominated by the flux part (similar to main **Fig 4c**).

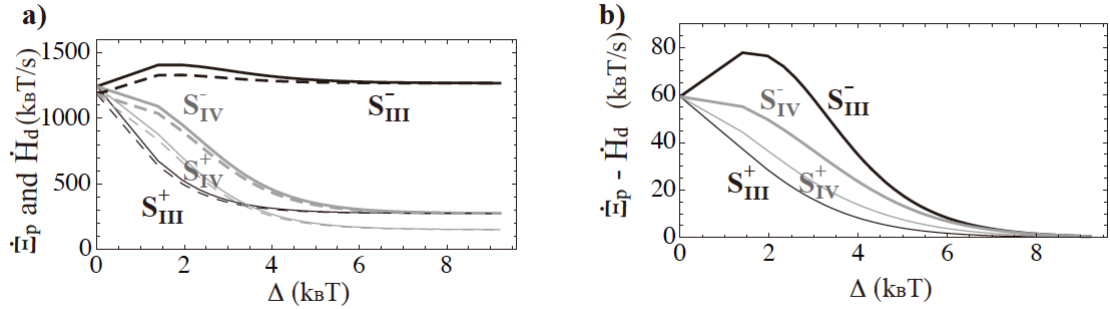

*Fig S5* Entropy production and heat dissipation during polymerase elongation. *(a)* The entropy production rate $\dot{\Xi}_P$ (solid lines) and the heat dissipation rate $\dot{H}_d$ (dashed lines) vs. $\Delta = k_BT\ln\eta$, for four elementary selections in the five-state kinetics. *(b)* The overall entropy change rate $\dot{\Xi}_P - \dot{H}_d$. The pre-selections $S_{III}^-$ and $S_{III}^+$ act the same (dark), while $S_{IV}^-$ and $S_{IV}^+$ act the same (gray).